# Neural Network Based Parameter Estimation Method for the Pareto/NBD Model

Shao-Ming XIE[1*]

[1]*Department of Business Administration, National Taiwan University, Taiwan, China (P.R.C)*

**Abstract:** Whether stochastic or parametric, the Pareto/NBD model can only be utilized for an in-sample prediction rather than an out-of-sample prediction. This research thus provides a neural network based extension of the Pareto/NBD model to estimate the out-of-sample parameters, which overrides the estimation burden and the application dilemma of the Pareto/NBD approach. The empirical results indicate that the Pareto/NBD model and neural network algorithms have similar predictability for identifying inactive customers. Even with a strong trend fitting on the customer count of each repeat purchase point, the Pareto/NBD model underestimates repeat purchases at both the individual and aggregate levels. Nonetheless, when embedding the likelihood function of the Pareto/NBD model into the loss function, the proposed parameter estimation method shows extraordinary predictability on repeat purchases at these two levels. Furthermore, the proposed neural network based method is highly efficient and resource-friendly and can be deployed in cloud computing to handle with big data analysis.
**Keywords:** Pareto/NBD model; Out-of-sample prediction; Neural network; Loss function; Likelihood function; Repeat purchase

---

[*] Corresponding author: D05741001@ntu.edu.tw

# 1. Introduction

The Pareto/NBD model developed by Schmittlein, Morrison, and Colombo (1987) (SMC hereafter) is a milestone within the group of Buy-Till-You-Die (BTYD) models, as it aims to formulate and forecast a customer's repeat buying behavior in a non-contractual setting. Many marketing researches have utilized this model, especially in the domain of customer relationship management (CRM hereafter), such as customer lifetime value (Gupta et al., 2006; Kumar & Reinartz, 2016) and retention estimation (Batislam, Denizel, & Filiztekin, 2007; Kamakura et al., 2005; Wübben & Wangenheim, 2008). For its attribution in customer base analysis, marketing academics have spent immerse efforts toward modifying the model and have provided fruitful variants under different implementation scenarios. However, some deficiencies in the BTYD model have made it sparsely known by people, especially for those who are not marketing background.

Initially, the key impediment comes from its modelling hypothesis. Whether it is stochastic or parametric, the BTYD model builds upon individual-level assumptions that can obtain the customized parameters that belong to a certain datapoint (or customer). Even if it able to gain a good estimation on the in-sample dataset (training dataset), the side effect is unable to help in the out-of-sample (testing dataset) prediction. In a big data context, this insufficiency brings an immeasurable estimation burden, especially for Markov Chain Monte Carlo (MCMC) estimation models. Secondly, except for the variant developed by Abe (2009), most BTYD models cannot easily incorporate the covariate effect into modelling. Thirdly, the Pareto/NBD model run under the maximum likelihood estimation (MLE) may face the numerical optimization problem of the complex likelihood function. If it is estimated by MCMC, then the maximum a posterior (MAP) approach is prone to overfitting, while the drawn parameters may overly fit and explain a single datapoint (Salakhutdinov & Mnih, 2008).

The deficiencies of the BTYD model impair the implementation opportunity in the business world, where machine learning has penetrated and established an irreplaceable status. Machine learning can discover patterns from the training dataset, parameterize the patterns via model optimization, and reuse the trained model for out-of-sample prediction (Murphy, 2012; Witten, Frank, Hall, & Pal, 2016). More importantly, machine learning can train the model by a subsample, and then the trained model can be harnessed to predict the out-of-sample dataset. In addition, machine learning can also include the covariate effect much easier than the BTYD model. Machine learning algorithm provides some techniques to prevent overfitting, such as the dropout function in the neural network algorithm, the pruning technique in the decision tree, and the slack variable in support vector machine. These advantages inspire the introduction of machine learning in this research to estimate the BTYD model's parameters. Among the numerous algorithms of machine learning, the neural network algorithm is the most flexible one as it can adjust the activation function, choose a different optimizer, and set a customized network structure. Moreover, it can present the complex non-linear relationships among input variables and output variables (Tu, 1996). With these benefits, this paper proposes the neural network algorithm based estimation method to estimate the BTYD model.

## 2. Literature Review

The Pareto/NBD model builds its assumptions on the transaction process and the dropout process, which are depicted by a Poisson distribution and an exponential distribution separately. Schmittlein et al. (1987) use two Gamma distributions to capture the heterogeneous transaction behavior across customers. Based on these assumptions, the Pareto/NBD model yields the alive probability and the expected repeat purchase that are employed in a lot of marketing research (Chan, Wu, & Xie, 2011; Reinartz & Venkatesan, 2008). Following their efforts, marketing researchers revised the Pareto/NBD model so as to meet the modelling needs of different business scenarios. Fader, Hardie, and Lee (2005) assume that the customer can drop out immediately after a transaction, which is depicted by the Beta-Geometric hypothesis rather than the Exponential-Gamma hypothesis. Thus, they propose the BG/NBD model. However, some customers drop out permanently after the first transaction, which cannot be captured by the BG/NBD model. Hence, Batislam et al. (2007) propose the Modified BG/NBD model to complement this missing part.

Based on the BG/NBD model, Fader, Hardie, and Shang (2010) raise a discrete version of the Pareto/NBD model, which is named the BG/BB model. This model replaces the Poisson-Gamma hypothesis of the Pareto/NBD model with the Bernoulli-Geometric hypothesis. Jerath, Fader, and Hardie (2011) propose a generalized version of the Pareto/NBD model, called the periodic death opportunity (PDO) model. It assumes that the customer periodically makes the defection decision, rather than immediately after a transaction or at any time after a transaction. The PDO model can approximate the Pareto/NBD model when the decision period is small and degenerates into the NBD model (Ehrenberg, 1972) when the decision period is large. These BTYD models are estimated by MLE, but some estimation burdens arise during the estimation, like solving the Gaussian Hypergeometric Function (Fader et al., 2005; Ma & Liu, 2007).

In order to avoid the estimation problem, Ma and Liu (2007) introduce MCMC into the parameter estimation. However, they leave the assumptions and the derivation of Pareto/NBD intact (Singh, Borle, & Jain, 2009), which is unable to fully take advantage of MCMC. Abe (2009) introduces the hierarchical bays framework and data augmentation into the Pareto/NBD model, relaxes the independent hypothesis between the transaction process and the dropout process with the multivariate normal distribution, and estimates the parameters by MCMC. His effort is known as the Pareto/NBD (Abe) model. More importantly, the Pareto/NBD (Abe) model can add the covariate effect into modelling much easier than other BTYD models. Platzer and Reutterer (2016) include transaction regularity into the BTYD model, which is described by the Gamma-Gamma hypothesis. Their effort is known as the Pareto/GGG model and is also estimated by MCMC.

Among the many selectable BTYD models, the Pareto/NBD model has shown its preeminent performance in many studies. Fader et al. (2005) find that the difference between the Pareto/NBD model and BG/NBD model is ignorable, and the former performs better in repeat transaction fitting. Through three empirical analyses, Abe (2009) also shows that the Pareto/NBD model has similar predictive performance as the Pareto/NBD (Abe) model. Aside from the flexible implementation of the PDO model, Jerath et al. (2011) suggest that managers can adopt the Pareto/NBD model for fitting future transactions. Moreover, Jasek, Vrana, Sperkova, Smutny, and Kobulsky (2019) conduct systematically comparisons between BTYD models in an online store dataset, concluding that the Pareto/NBD model has stable performance. Therefore, this research includes the Pareto/NBD model in the empirical applications and estimated it through the method proposed by Ma and Liu (2007), which can directly return individual-level parameters of the exponential distribution and Poisson distribution.

## 3. Empirical Method

### 3.1. Datasets

This paper exploits two datasets in the empirical analysis for comparison. The first dataset is the CDNOW dataset, which is the commonly used dataset in marketing research (Abe, 2009;

Fader & Hardie, 2001; Fader et al., 2005). It has 23,570 customers' purchase history up to the end of June 1997. The other dataset comes from an online clothing retailer in Taiwan (E-tailing hereafter), which records its customers' online transaction history. In order to provide a comparable customer base size to CDNOW, 24,000 customers are randomly sampled. The sampled E-tailing dataset has a total of 118,677 transactions at an average of NT$15,430 per customer. This research utilizes 60% of customers for training (the in-sample dataset) and the remainder for testing (the out-of-sample dataset).

Table 1. Data Description

|  | E-tailing | CDNOW |
| --- | --- | --- |
| Start Date | 2017-10-17 | 1997-01-01 |
| End Date | 2019-05-01 | 1998-06-30 |
| Number of Customers | 24,000 | 23,570 |
| Total Observations | 118,677 | 69,659 |
| Avg. Purchased CD Number per Customer | - | 7.12 |
| Avg. Sales per Customer | NT$ 15,430 | US$106.08 |

3.2. Data Preparation

Before the empirical application, this research first clarifies the data preparation procedure. The neural network algorithm and BTYD model are fed the same variables to conduct a fair comparison between the proposed estimation method and the BTYD model's estimation method. The following procedure explains how the data information is prepared for each dataset.

Step 1: The dataset is split into the in-sample dataset (or training dataset) and the out-of-sample dataset (or testing dataset). As Figure 1 shows, these subsamples are then split into the calibration period and the holdout period at time $T$, which is the mid-date of the dataset. The RFM summary (Recency, Frequency, Calibration Length) and the covariates are generated from the calibration period in the training dataset and the testing dataset with weekly data granularity.

Step 2: The generated variables of the training dataset in step 1 are the inputs for the BTYD models to obtain the individual-level parameters, $\lambda_i$ and $\mu_i$. These two estimated parameters are the output variables in the training process of the neural network algorithm. The generated variables of the testing dataset become input variables of the trained neural network algorithm to obtain the estimated parameters, $\hat{\lambda}_i$ and $\hat{\mu}_i$. Besides, the BTYD models estimate the inactive status and the repeat purchase in the out-of-sample dataset for comparison purpose.

Step 3: The estimated parameters from the neural network algorithm, $\hat{\lambda}_i$ and $\hat{\mu}_i$, are combined with the generated variables of the out-of-sample dataset to forecast the inactive status and the number of repeat purchases in the holdout period. These two estimated variables by the neural network algorithm are the quantities to compare with the quantities estimated by the BTYD models.

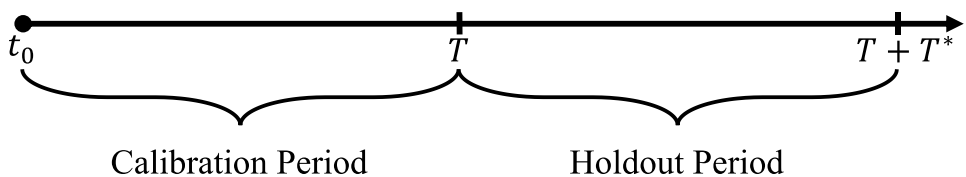

Figure 1. Data Preparation in the Pareto/NBD Model

3.3. Evaluation Metrics

(1) Accuracy for Inactive Status

Inactive status is the first quantity of concern in BTYD models. As a binary variable, accuracy is used to evaluate the correct classification. Higher accuracy demonstrates that the model has a more precise prediction.

$$\text{Accuracy} = \frac{TP+TN}{TP+FP+TN+FN} \quad (1)$$

Table 2. Confusion Matrix for Binary Classification

|  |  | Predicted Value | |
| --- | --- | --- | --- |
|  |  | Active | Inactive |
| Actual Value | Active | True Positive (TP) | False Negative (FN) |
|  | Inactive | False Positive (FP) | True Negative (TN) |

(2) Multi-class Accuracy and Prediction Consistency for the Purchase Number

In opposite to other research, this study extends the accuracy metric to evaluate the multi-class accuracy between the real purchase number and the predicted purchase number of the models during the holdout period. It evaluates the correctly prediction for the multi-class problem rather than the binary classification problem. In order to avoid the reading difficulty, the accuracy for the purchase number named as the multi-accuracy.

$$\text{Multi-accuracy} = \frac{1}{N}\sum_{i=1}^{N} I(y_i - \hat{y}_i), \ i = 1, 2, \ldots, N \quad (2)$$

Here, N is the customer number in the customer base; $y_i$ is the real number of purchases in the holdout period; $\hat{y}_i$ is the estimated number of purchases in the holdout period; and I(·) is the indicator function, which returns 1 when the model has a correct prediction. The greater multi-accuracy the model has, the better predictive power the model exhibits.

Moreover, the multi-accuracy can also be used to examine the prediction consistency between the BTYD model (BTYD) and the neural network algorithm (NNA). In the following, this research adopts the "Consistency(BTYD, NNA)" for this measurement purpose.

$$\text{Consistency(BTYD, NNA)} = \frac{1}{N}\sum_{i=1}^{N} I(\hat{y}_{BTYD,i} - \hat{y}_{NNA,i}), \ i = 1, 2, \ldots, N \quad (3)$$

Here, $\hat{y}_{BTYD,i}$ is the estimated purchase number by the BTYD model; and $\hat{y}_{NNA,i}$ is the estimated purchased number by the neural network algorithm. High consistency means that the BTYD model and the neural network algorithm shows consistency prediction at the individual level.

(3) Mean Absolute Error for the Purchase Number

Mean Absolute Error (MAE) is the metric to summarize and assess the prediction deviation of the number of purchases. The smaller the MAE value is, the better predictive power the model will have.

$$\text{MAE} = \frac{1}{N}\sum_{i=1}^{N} |y_i - \hat{y}_i|, \ i = 1, 2, \ldots, N \quad (4)$$

3.4. Proposed Modelling Schema

3.4.1. Loss Function Setting Up

As a supervised learning algorithm, the neural network algorithm needs output variables in the training part to solve the cold start problem. Hence, this research uses BTYD models to obtain the parameters ($\lambda$ and $\mu$) for each datapoint as the output variables in the neural network training. This research introduces two kinds of loss function approaches: one is without a likelihood function and the other is with a likelihood function. Eight corresponding experiments with different loss functions are then exploited to estimate the parameters.

The without likelihood function approach uses two commonly adopted loss functions: Mean Absolute Error (MAE hereafter) and Mean Square Error (MSE hereafter). Both of them are as the comparison base to verify the advantage of the loss function with the embedded likelihood function of the corresponding BTYD model. These two neural network algorithms are denoted as NNA (MAE) and NNA (MSE).

The other approach embeds the likelihood function of a specific BTYD model into the loss function. This approach provides two embedding strategies: one is based on the likelihood function, which consists of the likelihood function (Likelihood hereafter), the likelihood function with MSE (Likelihood + MSE hereafter), and the likelihood function with MAE (Likelihood + MAE hereafter); the second one is an embedding strategy based on the likelihood function ratio (Likelihood Ratio hereafter), the likelihood function ratio with MSE (hereafter Likelihood Ratio + MSE), and the likelihood function ratio with MAE (Likelihood Ratio + MAE hereafter). The corresponding neural network algorithms with these loss functions are denoted as NNA (Likelihood), NNA (Likelihood + MAE), NNA (Likelihood + MSE), NNA (Likelihood Ratio), NNA (Likelihood Ratio + MAE), and NNA (Likelihood Ratio + MSE). The individual-level likelihood function of the Pareto/NBD model is (Fader & Hardie, 2005; Ma & Liu, 2007):

$$L(x, t_x, T|\lambda, \mu) = \frac{\lambda^x}{\lambda+\mu}(\mu e^{-(\lambda+\mu)t_x} + \lambda e^{-(\lambda+\mu)T}) \qquad (5)$$

Here, $x$ denotes repeat transactions in the calibration period; $t_x$ denotes recency, which is the time between the first-ever transaction and the last transaction; $T$ denotes calibration length; and $\lambda$ and $\mu$ are the parameters of Poisson distribution and the exponential distribution.

Compared to MAE and MSE, the neural network algorithm with the likelihood function must embed Recency, Frequency, and Calibration Length in the likelihood-based loss function. Figure 2 shows the data flow of the likelihood based neural network algorithm in a non-covariate scenario. Here, $i$ is the datapoint (or customer ID); Calibration Length$_i$, Recency$_i$, and Frequency$_i$ are the input variables; and $\lambda_i$ and $\mu_i$ are the corresponding outputs. The predicted $\hat{\mu}_i$ and $\hat{\lambda}_i$ with the inputs are used to calculate $Likelihood_i$. Then, the sum of the training data's log likelihood multiply by -1 as the loss function to optimize the network.

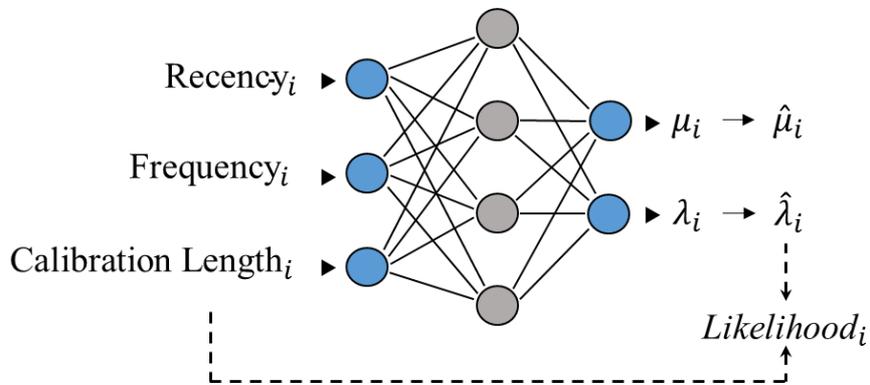

Figure 2. Likelihood Based Neural Network Algorithm

A neural network algorithm can revise the loss function much easier so as to incorporate more information. The Likelihood + MSE function is $Likelihood_i + (\mu_i - \hat{\mu}_i)^2 + (\lambda_i - \hat{\lambda}_i)^2$, and the Likelihood + MAE function is $Likelihood_i + |\mu_i - \hat{\mu}_i| + |\lambda_i - \hat{\lambda}_i|$, which are designed to maximize the likelihood when the predicted parameters approximate the parameters of the Pareto/NBD model as close as possible. The Likelihood Ratio is $Likelihood_i(\hat{\mu}_i, \hat{\lambda}_i)/Likelihood_i(\mu_i, \lambda_i)$, which aims to approach the likelihood value of the Pareto/NBD model. Same as that of Likelihood, Likelihood Ratio with $(\mu_i - \hat{\mu}_i)^2 + (\lambda_i - \hat{\lambda}_i)^2$ or $|\mu_i - \hat{\mu}_i| + |\lambda_i - \hat{\lambda}_i|$ is denoted as Likelihood Ratio + MSE and Likelihood Ratio + MAE to estimate the parameters.

3.4.2. Neural Network Structure

Through multiple events of trial-and-error, the utilized neural network structure in this research comes with two hidden layers, 20 neurons in each layer, and the Sigmoid activation function in each neuron. In order to solve the overfitting problem, this research adopts a 20% dropout probability in the training.

3.4.3. Analytical Workflow

This research consists of three stages in each empirical analysis.

(1) The first stage obtains the parameters ($\mu_i$ and $\lambda_i$), the predicted inactive status, and the estimated number of purchases by the Pareto/NBD model.

(2) The second stage trains the neural network algorithm with MAE, MSE, or the likelihood function. The trained models are then used to predict the inactive status and repeat purchase in the holdout period.

(3) The third stage calculates the evaluation metrics by the true and estimated inactive status and repeat purchase. Conclusions are then drawn from the comparison.

## 4. Empirical Applications

This section follows the previous analytical procedure to conduct the empirical analysis with two real-world datasets. The empirical results are then used to define the advantage of the proposed parameter estimation methods.

4.1. CDNOW Dataset

Figure 3 shows the neural network algorithms present a similar prediction at zero repeat transaction prediction, but overestimate the customer numbers for non-zero repeat transactions. The prediction made by Pareto/NBD at zero repeat transactions is relatively higher than the true customer numbers, and it has smaller customer number fits in each non-zero repeat transaction.

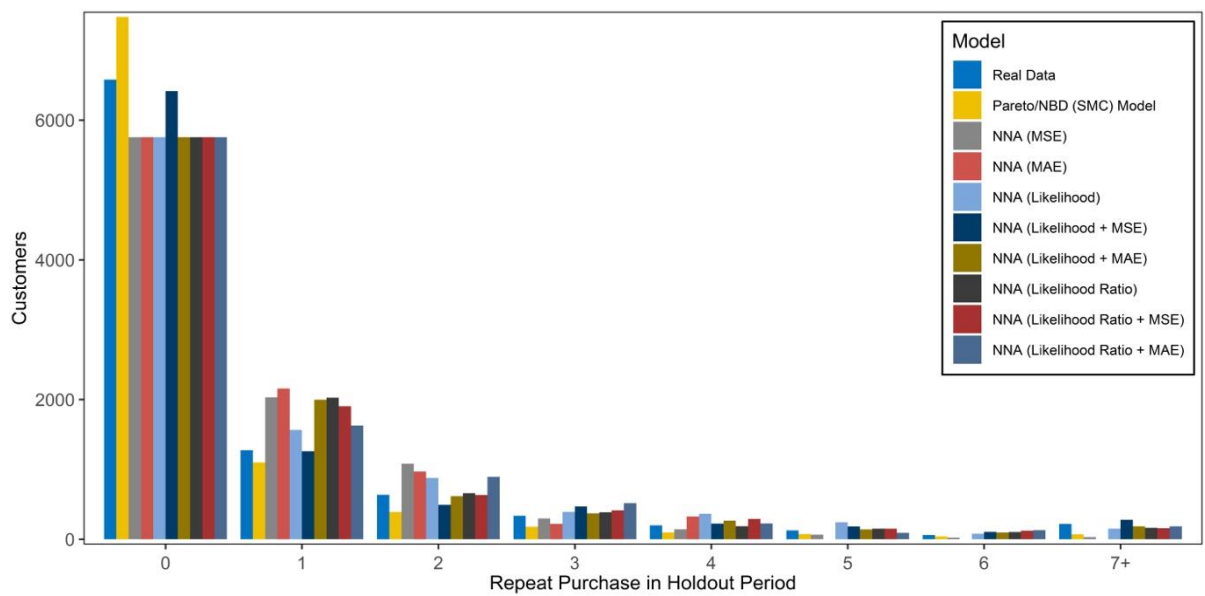

Figure 3. Predicted Versus Actual Customer Number of Repeat Purchases (CDNOW)

This study looks further into the repeat purchases in Figure 3. Table 3 demonstrates that the Pareto/NBD model offers a conservative estimation on repeat transactions (Jain & Singh, 2002; Ma & Büschken, 2011). It predicts 7,479 customers will conduct no repeat transaction in the holdout period, while the remaining customer number predictions at each non-zero repeat purchase number are smaller than the true customer numbers. However, the NNA (Likelihood + MSE) performs best in customer number fitting, as it has the same trend or proportional prediction to the real dataset. Others neural network algorithms have similar predictability on customer number fitting. But NNA (MAE), NNA (MSE), and NNA (Likelihood Ratio + MAE) do not fit well at small purchases number.

Table 3. Predicted Versus Actual Customer Number of Repeat Purchases (CDNOW)

| Repeat Purchase Number | 0 | 1 | 2 | 3 | 4 | 5 | 6 | 7+ |
|---|---|---|---|---|---|---|---|---|
| Pareto/NBD model | 7,479 | 1,099 | 391 | 178 | 97 | 75 | 39 | 70 |
| Real Data | 6,582 | 1,274 | 635 | 336 | 197 | 126 | 60 | 218 |
| NNA (MSE) | 5,757 | 2,032 | 1,081 | 298 | 141 | 66 | 23 | 30 |
| NNA (MAE) | 5,757 | 2,158 | 969 | 219 | 325 | 0 | 0 | 0 |
| NNA (Likelihood) | 5,757 | 1,564 | 878 | 392 | 364 | 244 | 79 | 150 |
| NNA (Likelihood + MSE) | 6,416 | 1,260 | 492 | 469 | 224 | 182 | 105 | 280 |
| NNA (Likelihood + MAE) | 5,757 | 1,997 | 615 | 371 | 268 | 140 | 96 | 184 |
| NNA (Likelihood Ratio) | 5,757 | 2,025 | 658 | 386 | 185 | 151 | 103 | 163 |
| NNA (Likelihood Ratio + MSE) | 5,757 | 1,904 | 632 | 414 | 292 | 151 | 121 | 157 |
| NNA (Likelihood Ratio + MAE) | 5,757 | 1,629 | 894 | 516 | 227 | 91 | 130 | 184 |

The customer frequency prediction in Figure 3 is an aggregate level summarization. The conclusion drawn from Table 3 by eye-balling is a visualized conclusion, which cannot be evidence to determine the optimal model. Thus, Table 4 shows the evaluation criteria in the optimal modelling candidate selection.

First, the Pareto/NBD model and all neural network algorithms have similar inactive status predictions. Almost 77% of customers turn into an inactive status in the holdout period.

Second, the aggregate repeat purchases made by 9,428 customers (40% of the out-of-sample dataset) in the holdout period is 7,634. Echoing Figure 3 and Table 3, even the Pareto/NBD model shows a relative good trend fitting of the customer frequency, but it has the worst purchase number prediction at 4,105 in total. This underestimation of the Pareto/NBD model on repeat transactions is due to the conservative estimation, especially in zero repeat purchase fitting. NNA (Likelihood Ratio) has the most accurate purchase estimation at the aggregate level, which shows a minimum prediction deviation in the holdout period.

Third, NNA (Likelihood + MAE) and NNA (Likelihood Ratio + MSE) have suboptimum forecasting. NNA (Likelihood + MSE), which presents a good trend fit in Table 3 to true customer distribution, loses its advantage on the aggregate purchase number prediction, but it still performs better than the Pareto/NBD model, NNA (MAE), and NNA (MSE). The comparison in fitting performance delivers a critical insight into modelling, whereby the customer distribution on repeat purchases does not depict the true future repeat purchases at the individual and aggregate levels.

Table 4. Evaluation Metrics (CDNOW)

| Model | Inactive Accuracy | Multi-accuracy | MAE | Consistency (NNA, SMC) | Number of Purchases |
|---|---|---|---|---|---|
| **Pareto/NBD model** | 76.56% | 67.14% | 0.690 | - | 4,105 |
| **NNA (MSE)** | 76.63% | 57.98% | 0.769 | 71.34% | 6,446 |
| **NNA (MAE)** | 76.87% | 58.04% | 0.772 | 71.61% | 6,053 |
| **NNA (Likelihood)** | 76.27% | 56.66% | 0.904 | 67.11% | 8,975 |
| **NNA (Likelihood + MSE)** | 76.68% | 61.09% | 0.896 | 73.76% | 8,526 |
| **NNA (Likelihood + MAE)** | 76.54% | 57.20% | 0.862 | 66.54% | 8,492 |
| **NNA (Likelihood Ratio)** | 76.44% | 57.27% | 0.840 | 67.12% | 7,925 |
| **NNA (Likelihood Ratio + MSE)** | 76.42% | 57.09% | 0.859 | 66.66% | 8,438 |
| **NNA (Likelihood Ratio + MAE)** | 76.55% | 56.84% | 0.878 | 66.01% | 8,809 |

With the evidence in Figure 3 and Table 3, the evaluation metrics can be used to discover the model's predictive focus and strength. The greatest prediction weight of the Pareto/NBD model is on zero repeat purchase, whereas other non-zero repeat purchases receive relatively insufficient focus. This contributes to its underestimation of total repeat purchases in the holdout period. The multi-accuracy also verifies its main predictive focus is on zero repeat purchase, where the Pareto/NBD model has the best exact prediction at individual-level repeat purchase fitting among all utilized models. In spite of its over-prediction in the number of purchases, NNA (Likelihood + MSE) exhibits a multi-accuracy over 60%. It has the highest predictive consistency to the Pareto/NBD model at individual-level repeat transaction prediction among the neural network algorithms, but shows severe deviation from the real transaction number. Tracing back to the customer distribution in Table 3, this over-estimation indicates that it puts greater predictive weight on non-zero repeat purchase prediction. NNA (Likelihood Ratio) has only 67.12% consistency with the Pareto/NBD model and a lower multi-accuracy value. Compared to the 61.09% multi-accuracy of NNA (Likelihood + MSE), it has an acceptable individual repeat purchase estimation at 57.27% multi-accuracy. In conclusion, this study regards NNA (Likelihood Ratio) as the best model for approximating real data in the without-covariate setting. Aside from it, NNA (Likelihood Ratio + MSE), NNA (Likelihood + MSE), and NNA (Likelihood + MAE) also have the modelling opportunity in the business analysis.

Table 5. Correlations Between Multi-accuracy, MAE, Number of Purchases, and Consistency (CDNOW)

| Correlation | Multi-accuracy | MAE | Number of Purchases | Consistency |
|---|---|---|---|---|
| **Multi-accuracy** | 1.0000 | | | |
| **MAE** | -0.6663 | 1.0000 | | |
| **Number of Purchase** | -0.7496 | 0.9846 | 1.0000 | |
| **Consistency** | 0.8668 | -0.3681 | -0.5501 | 1.0000 |

In Table 5 the correlation between the multi-accuracy and MAE is -0.6663, which demonstrates that correctly prediction does not contribute to a lower MAE. The correlation between the multi-accuracy and the number of purchases is -0.7496, meaning that underestimation on the purchase number brings out a higher multi-accuracy. The correlation of 0.9846 between MAE and the total number of purchases demonstrates that an underestimation of total transactions by the Pareto/NBD model has a lower MAE - that is, the neural network algorithms with a relatively higher MAE perform better in total transaction number prediction. Alongside this, the model pours intensive predictive weights on zero repeat transactions, and thus there will be greater multi-accuracy. Furthermore, the inconsistency between the neural network model and the Pareto/NBD model suggests that a dissimilar prediction by the neural network algorithm can generate a better fitting at individual repeat transactions, as proven in customer distribution and total repeat purchases.

4.2. E-tailing Dataset

The E-tailing dataset has a dispersed customer distribution on each repeat purchase number, and the following summarization is counted until 17+.

NNA (MAE) or NNA (MSE) has the worst performance on customer distribution fitting over each repeat purchase point. They both underestimate the customer number on the zero transaction point and overestimate the customer number on one and two repeat transactions. Different with the CDNOW dataset, the Pareto/NBD model presents unsatisfied customer number fitting on each repeat transaction point, but the neural network algorithms with likelihood function have a good fit in customer counts on each repeat purchase point.

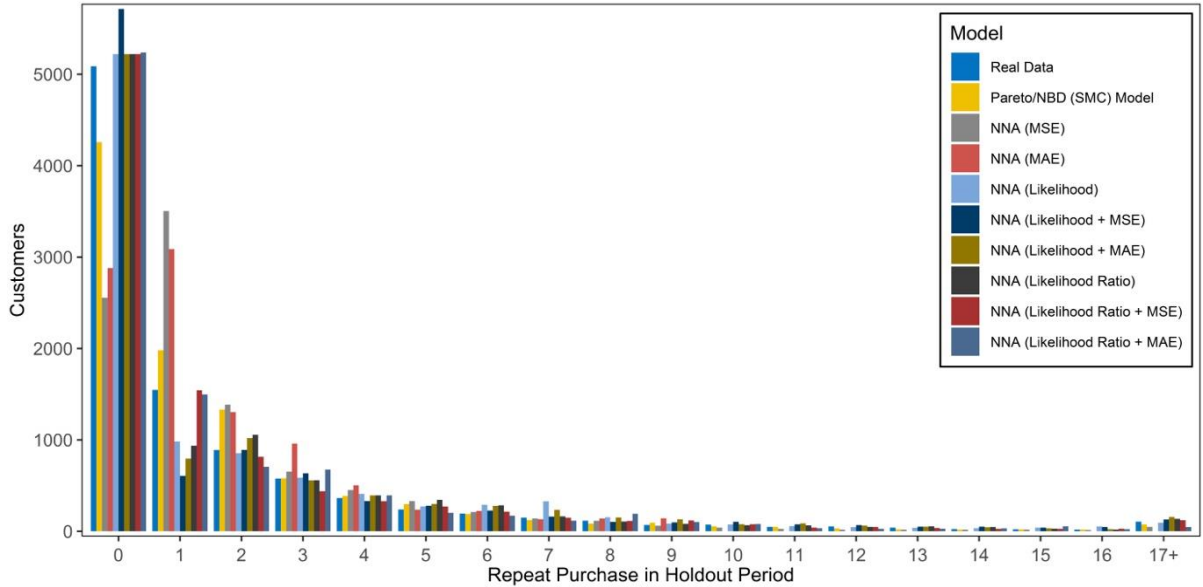

Figure 4. Predicted Versus Actual Customer Number of Repeat Purchases (E-tailing)

Table 6 reports the customer count behind Figure 4, where NNA (Likelihood Ratio + MAE) and NNA (Likelihood Ratio + MSE) have the best approximation to the true customer distribution on each repeat transaction point. In addition, NNA (MAE) and NNA (MSE) cannot be modelling candidates due to abnormal customer count fitting. Even if the Pareto/NBD model has good customer distribution approximation, the severely incorrect fitting on low repeat purchase points shows inaccurate predictability. Just like in the previous dataset, the customer distribution may emit erroneous evidence in the modelling decision. The evaluation metrics will be introduced again to find the optimum modelling candidate.

Table 6. Predicted Versus Actual Customer Number of Repeat Purchases (E-tailing)

| Repeat Purchase Number | 0 | 1 | 2 | 3 | 4 | 5 | 6 | 7 | 8 | 9 | 10 | 11 | 12 | 13 | 14 | 15 | 16 | 17+ |
|---|---|---|---|---|---|---|---|---|---|---|---|---|---|---|---|---|---|---|
| Pareto/NBD Model | 4,257 | 1,980 | 1,331 | 579 | 385 | 297 | 191 | 122 | 83 | 93 | 54 | 48 | 30 | 21 | 18 | 20 | 17 | 74 |
| Real Data | 5,087 | 1,545 | 889 | 577 | 363 | 238 | 194 | 147 | 115 | 69 | 72 | 48 | 52 | 39 | 23 | 21 | 17 | 104 |
| NNA (MSE) | 2,554 | 3,504 | 1,385 | 653 | 451 | 330 | 211 | 139 | 114 | 61 | 40 | 26 | 17 | 17 | 17 | 19 | 14 | 48 |
| NNA (MAE) | 2,879 | 3,087 | 1,304 | 959 | 502 | 236 | 223 | 131 | 138 | 141 | 0 | 0 | 0 | 0 | 0 | 0 | 0 | 0 |
| NNA (Likelihood) | 5,221 | 982 | 852 | 586 | 409 | 272 | 289 | 327 | 152 | 82 | 74 | 56 | 44 | 37 | 33 | 39 | 52 | 93 |
| NNA (Likelihood + MSE) | 5,715 | 606 | 891 | 634 | 330 | 279 | 227 | 159 | 102 | 97 | 103 | 75 | 67 | 50 | 50 | 40 | 47 | 128 |
| NNA (Likelihood + MAE) | 5,221 | 794 | 1,019 | 557 | 392 | 299 | 277 | 235 | 150 | 130 | 77 | 86 | 62 | 50 | 43 | 31 | 22 | 155 |
| NNA (Likelihood Ratio) | 5,221 | 936 | 1,055 | 559 | 393 | 344 | 283 | 162 | 106 | 80 | 66 | 64 | 48 | 54 | 47 | 28 | 19 | 135 |
| NNA (Likelihood Ratio + MSE) | 5,221 | 1,540 | 814 | 438 | 329 | 270 | 215 | 146 | 113 | 117 | 77 | 40 | 46 | 35 | 26 | 26 | 27 | 120 |
| NNA (Likelihood Ratio + MAE) | 5,238 | 1,496 | 705 | 674 | 392 | 202 | 170 | 115 | 192 | 100 | 79 | 33 | 26 | 27 | 31 | 53 | 22 | 45 |

First, all the neural network algorithms are sophisticated at identifying inactive customers during the holdout period, whereas the Pareto/NBD model shows unexpected predictive accuracy. The best forecasting can achieve almost 70% correct classification.

Second, the total transaction number in the holdout period made by 9,600 customers is 17,159. It seems the Pareto/NBD model has an accurate total repeat purchase estimation, but has a low multi-accuracy in Table 7 and a poor customer count distribution in Table 6. Conversely, NNA (Likelihood Ratio + MSE) has the best aggregate-level repeat purchase estimation and good multi-accuracy and MAE among the neural network algorithms.

Third, the neural network algorithms with likelihood function and NNA (Likelihood Ratio) overestimate the total repeat transactions, which leads to high MAE. However, these models have better multi-accuracy and MAE than the Pareto/NBD model. NNA (Likelihood Ratio + MAE) underestimates total repeat transactions in the holdout period. These discoveries denote that the embedded likelihood function in loss function is conducive to training and obtaining a good neural network structure, especially when MSE is included. Generally speaking, NNA (Likelihood Ratio + MSE) is the optimal modelling candidate in this dataset, because it has the best customer distribution and a good fit for individual- and aggregate-level repeat transactions.

Table 7. Evaluation Metrics (E-tailing)

| Model | Inactive Accuracy | Multi-accuracy | MAE | Consistency (NNA, SMC) | Number of Purchases |
|---|---|---|---|---|---|
| **Pareto/NBD model** | 57.70% | 37.85% | 1.573 | - | 17,066 |
| **NNA (MSE)** | 65.48% | 29.26% | 1.576 | 53.92% | 18,140 |
| **NNA (MAE)** | 66.06% | 30.75% | 1.589 | 52.31% | 16,388 |
| **NNA (Likelihood)** | 68.44% | 42.71% | 1.698 | 47.15% | 19,487 |
| **NNA (Likelihood + MSE)** | 68.77% | 45.69% | 1.687 | 49.80% | 18,774 |
| **NNA (Likelihood + MAE)** | 68.38% | 42.43% | 1.793 | 43.43% | 21,291 |
| **NNA (Likelihood Ratio)** | 69.08% | 43.00% | 1.669 | 46.40% | 19,394 |
| **NNA (Likelihood Ratio + MSE)** | 65.70% | 44.11% | 1.596 | 55.76% | 17,418 |
| **NNA (Likelihood Ratio + MAE)** | 66.05% | 44.38% | 1.516 | 56.14% | 15,982 |

In Table 8 the number of purchases and MAE maintain a high correlation like that in the CDNOW dataset, suggesting a positive and stable relationship between these two metrics. On the contrary, the E-tailing dataset shows weak and positive correlations between the multi-accuracy and MAE or the number of purchases. This implies that the multi-accuracy has a little impact on MAE or the number of purchases. Consequently, high consistency with the Pareto/NBD model brings about dissatisfaction in the multi-accuracy on individual-level repeat transaction forecasting. Additionally, the neural network algorithms with likelihood function perform better than the Pareto/NBD model when a correctly prediction presents at individual-level repeat purchase prediction.

Table 8. Correlation Between Multi-accuracy, MAE, and Number of Purchases (E-tailing)

| Correlation | Multi-accuracy | MAE | Number of Purchases | Consistency |
|---|---|---|---|---|
| **Multi-accuracy** | 1.0000 | | | |
| **MAE** | 0.3528 | 1.0000 | | |
| **Number of Purchase** | 0.2994 | 0.9443 | 1.0000 | |
| **Consistency** | -0.2369 | -0.9309 | -0.9005 | 1.0000 |

4.3. Discussion and Management Insights

4.3.1. Discussion

The empirical results in general demonstrate that the proposed parameter estimation method with the neural network algorithm shows extraordinary performance over the Pareto/NBD model at inactive customer identification and repeat purchase estimation (at both the individual and aggregate levels). Even if the label in the training process is from the Pareto/NBD model, the neural network algorithm with likelihood function as/in loss function can generate better parameters for inactive status estimation and repeat purchase estimation. There are several useful implications gained from the above empirical analysis.

First, all the included models have similar prediction accuracy in identifying inactive customers. The likelihood function is ignorable in this quantity estimation, because the conventional neural network algorithm with MAE or MSE as loss function can satisfy analytical needs and management needs.

Second, the Pareto/NBD model offers insufficient estimation of aggregate-level repeat purchases, which may derives from its "buy-till-die" assumption (Jain & Singh, 2002; Ma & Büschken, 2011). The seemingly strong fitting of the Pareto/NBD model in customer frequency on each repeat purchase point disguises its conservative estimations. The Pareto/NBD model weights more predictive power on zero repeat transaction fitting, which leads to insufficient non-zero repeat purchase fitting. On the contrary, the neural network algorithms with likelihood function emit better individual- and aggregate-level estimations.

Third, NNA (Likelihood) is not the best modelling choice among the four comparison settings. If the loss function adds the constraint of MAE, MSE, or the likelihood function of the Pareto/NBD model, then predominant improvement in fitting can be obtained - that is, the included MAE or MSE constrains the estimated parameters to be as close to the parameters of the Pareto/NBD model. The likelihood value with the parameters estimated by the neural network algorithm can also approximate the likelihood value of the Pareto/NBD model. On the other hand, the included likelihood function is conducive to the neural network algorithm obtaining a better error for network optimization than the model with only MAE or MSE.

Fourth, the negative and high consistency between the Pareto/NBD model and the neural network algorithm indicates that the neural network algorithm outperforms in individual-level repeat purchase estimation. The extraordinary power of the neural network algorithm derives from the average predictive weight on each repeat purchase point rather than on the zero repeat purchase point.

4.3.2. Managerial Insights

The Pareto/NBD model has been proven to exhibit outstanding implementation in CRM. However, model assumptions constrain the spillover prediction in an out-of-sample dataset. By contrast, machine learning has permeated and dominated most industries in business practice. This paper thus provides an extension to estimate the parameters of the Pareto/NBD model, offering a better performance than the Pareto/NBD model, NNA (MAE), and NNA (MSE). The proposed estimation method shows implementation opportunity in real business applications.

First, the neural network algorithm with likelihood function can estimate a more precise repeat purchase number at the individual and aggregate levels during the holdout period. A precise prediction of purchase number is helpful for inventory management, as it saves on inventory cost and supports better inventory planning. In addition, it is useful for one-to-one marketing, which can identify a customer who are going to make a repeat purchase in the future and how many transactions he/she will conduct.

Second, the neural network algorithm with likelihood function can be deployed in cloud computing to conduct individual-level prediction on big data. Under the estimation method of MCMC, the estimation and prediction by the Pareto/NBD model are very resource-consuming and time-consuming (Bijmolt et al., 2010). For the implementation of the CDNOW dataset in the without-covariate setting, this research employs Mac Pro 2017 with a 2.5 GHz Intel Core i7 processor and 16 GB memory. The total running time of parameter estimation and quantity forecasting of the out-of-sample dataset is 3 minutes 57 seconds. However, NNA (Likelihood) in the training and testing process only took 21 seconds. This is the advantage of the proposed estimation model in business practice.

Third, the BTYD model is a new modelling candidate, besides machine learning, for customer-based analysis. The dominating status of machine learning cannot be changed or even be challenged. However, the proposed parameter estimation method overrides the BTYD model from estimation dilemma and is easily implemented in real business practices. Moreover, the formulations of the BTYD model are left intact, and the only effort is to find an appropriate neural network structure for the specific implementation scenario.

## 5. Conclusions

This research proposes a new estimation method to estimate the individual-level parameters of the Pareto/NBD model by a neural network algorithm. The unique difference in the proposed neural network algorithm compared to a typical neural network algorithm is the likelihood function of the Pareto/NBD model is embedded in the loss function. This revision presents high efficiency parameter estimation and contributes to a better prediction during the inactive status and on repeat purchases in the holdout period. Moreover, it can embed more loss information with the likelihood function so as to generate a better error for the backpropagation process in network optimization. The neural network algorithm is more flexible and efficient than MCMC in parameter estimation, as it can include the covariate effect and revise the network structure much easier. This research also conducts the comparison between the Pareto/NBD (Abe) model and the proposed parameter estimation model under the with-covariate setting (find at the supplementary material), which has the similar results as the above without-covariate setting. More importantly, the dropout function in the neural network algorithm can avoid the overfitting problem that may exist in MCMC. Finally, the neural network model is a non-linear model and thus can illustrate the non-linear relationships among variables into feature engineering. This is a significant improvement compared to Abe (2009), who only utilizes a linear relationship in the multivariate normal distribution.

Even though the proposed estimation model has fruitful benefits and significant improvements, it still has some deficiencies that need to be resolved. First, as a model with a "black box" nature, the neural network algorithm cannot extract the interpretable relationship between Recency, Frequency, Calibration length, and the added covariates. It cannot obtain a numerical interpretable relationship among covariates as can that in Abe (2009). Additionally, the loss function in the optimum neural network algorithm in this research is not the Likelihood, even this research has tried different network structures and different activation functions in the trial-and-error experiments. In order to achieve or surpass the predictability of the BTYD model, the loss function with likelihood function should include an additional constraint, like MAE or MSE. This deficiency should thus receive more efforts to separate the contribution of the constraint in optimizing the neural network algorithm.

Through the findings herein, there are some future research directions that can be taken. Initially, the individual-level repeat purchase is crucial for the customer lifetime value literature (Borle, Singh, & Jain, 2008; Glady, Baesens, & Croux, 2009; Reinartz & Kumar, 2000). One foreseeable future work is to incorporate the Gamma-Gamma monetary model (Fader & Hardie, 2013) to estimate a more accurate individual lifetime value. In addition, the next visiting time of the customer is the quantity of most concern in academic research and in the business world. To understand this problem, one should model the inter-transaction time into the analysis, and thus the Pareto/GGG model can be adopted. Moreover, a researcher can utilize the sequence data directly for the estimation, but not feed the data into the BTYD model. Last but not least, this research adopts the frequently-used loss functions, MAE and MSE, into the loss function. There are other loss functions that can be added into the loss function, such as Kullback-Leibler divergence. This points to another avenue to take for related investigations.

## References

Abe, M. (2009). "Counting your customers" One by One: A Hierarchical Bayes Extension to the Pareto/NBD Model. *Marketing Science, 28*(3), 541-553.

Batislam, E. P., Denizel, M., & Filiztekin, A. (2007). Empirical Validation and Comparison of Models for Customer Base Analysis. *International Journal of Research in Marketing, 24*(3), 201-209.

Bijmolt, T. H., Leeflang, P. S., Block, F., Eisenbeiss, M., Hardie, B. G., Lemmens, A., & Saffert, P. (2010). Analytics for customer engagement. *Journal of Service Research, 13*(3), 341-356.

Borle, S., Singh, S. S., & Jain, D. C. (2008). Customer Lifetime Value Measurement. *Management Science, 54*(1), 100-112.

Chan, T. Y., Wu, C., & Xie, Y. (2011). Measuring the Lifetime Value of Customers Acquired from Google Search Advertising. *Marketing Science, 30*(5), 837-850.

Ehrenberg, A. S. C. (1972). *Repeat-buying: Theory and Applications*: North-Holland Amsterdam.

Fader, P. S., & Hardie, B. G. (2001). Forecasting Repeat Sales at CDNOW: A Case Study. *Interfaces, 31*(3_supplement), S94-S107.

Fader, P. S., & Hardie, B. G. (2005, 01/01). A note on deriving the Pareto/NBD model and related expressions. Retrieved from http://brucehardie.com/notes/009/

Fader, P. S., & Hardie, B. G. (2013). The Gamma-Gamma model of monetary value. Retrieved from http://www.brucehardie.com/notes/025/

Fader, P. S., Hardie, B. G., & Lee, K. L. (2005). "Counting your customers" the Easy Way: An Alternative to the Pareto/NBD Model. *Marketing Science, 24*(2), 275-284.

Fader, P. S., Hardie, B. G., & Shang, J. (2010). Customer-base Analysis in a Discrete-time Noncontractual Setting. *Marketing Science, 29*(6), 1086-1108.

Glady, N., Baesens, B., & Croux, C. (2009). A Modified Pareto/NBD Approach for Predicting Customer Lifetime Value. *Expert Systems with Applications, 36*(2), 2062-2071.

Gupta, S., Hanssens, D., Hardie, B., Kahn, W., Kumar, V., Lin, N., . . . Sriram, S. (2006). Modeling Customer Lifetime Value. *Journal of Service Research, 9*(2), 139-155.

Jain, D., & Singh, S. S. (2002). Customer Lifetime Value Research in Marketing: A Review and Future Directions. *Journal of Interactive Marketing, 16*(2), 34.

Jasek, P., Vrana, L., Sperkova, L., Smutny, Z., & Kobulsky, M. (2019). Comparative Analysis of Selected Probabilistic Customer Lifetime Value Models in Online Shopping. *Journal of Business Economics and Management, 20*(3), 398-423.

Jerath, K., Fader, P. S., & Hardie, B. G. (2011). New Perspectives on Customer "Death" Using a Generalization of the Pareto/NBD Model. *Marketing Science, 30*(5), 866-880.

Kamakura, W., Mela, C. F., Ansari, A., Bodapati, A., Fader, P., Iyengar, R., . . . Verhoef, P. C. (2005). Choice Models and Customer Relationship Management. *Marketing Letters, 16*(3-4), 279-291.

Kumar, V., & Reinartz, W. (2016). Creating Enduring Customer Value. *Journal of Marketing, 80*(6), 36-68.

Ma, S.-H., & Büschken, J. (2011). Counting your customers from an "always a share" perspective. *Marketing Letters, 22*(3), 243-257. doi:10.1007/s11002-010-9123-0

Ma, S.-H., & Liu, J.-L. (2007). *The MCMC Approach for Solving the Pareto/NBD Model and Possible Extensions.* Paper presented at the Third International Conference on Natural Computation (ICNC 2007).

Murphy, K. P. (2012). *Machine Learning: A Probabilistic Perspective*: MIT Press.

Platzer, M., & Reutterer, T. (2016). Ticking Away the Moments: Timing Regularity Helps to Better Predict Customer Activity. *Marketing Science, 35*(5), 779-799.

Reinartz, W. J., & Kumar, V. (2000). On the profitability of long-life customers in a noncontractual setting: An empirical investigation and implications for marketing. *Journal of Marketing, 64*(4), 17-35.

Reinartz, W. J., & Venkatesan, R. (2008). Decision Models for Customer Relationship Management (CRM). In *Handbook of Marketing Decision Models* (pp. 291-326): Springer.

Salakhutdinov, R., & Mnih, A. (2008). *Bayesian Probabilistic Matrix Factorization using Markov Chain Monte Carlo.* Paper presented at the Proceedings of the 25th International Conference on Machine Learning.

Schmittlein, D. C., Morrison, D. G., & Colombo, R. (1987). Counting Your Customers: Who-are They and What Will They Do Next? *Management Science, 33*(1), 1-24.

Singh, S. S., Borle, S., & Jain, D. C. (2009). A Generalized Framework for Estimating Customer Lifetime Value When Customer Lifetimes Are Not Observed. *Quantitative Marketing and Economics, 7*(2), 181-205.

Tu, J. V. (1996). Advantages and Disadvantages of Using Artificial Neural Networks versus Logistic Regression for Predicting Medical Outcomes. *Journal of Clinical Epidemiology, 49*(11), 1225-1231.

Wübben, M., & Wangenheim, F. v. (2008). Instant customer base analysis: Managerial heuristics often "get it right". *Journal of Marketing, 72*(3), 82-93.

Witten, I. H., Frank, E., Hall, M. A., & Pal, C. J. (2016). *Data Mining: Practical Machine Learning Tools and Techniques*: Morgan Kaufmann.

# Appendix. Modelling with the Covariate Effect

1. CDNOW dataset

By incorporating the CD number and Sales as covariates into the BTYD model, this study uses the Pareto/NBD (Abe) model to conduct the empirical analysis to compare predictability with the neural network algorithms. Figure 5 shows the similar customer distribution over each repeat purchase point as that in the non-covariate setting in Figure 3.

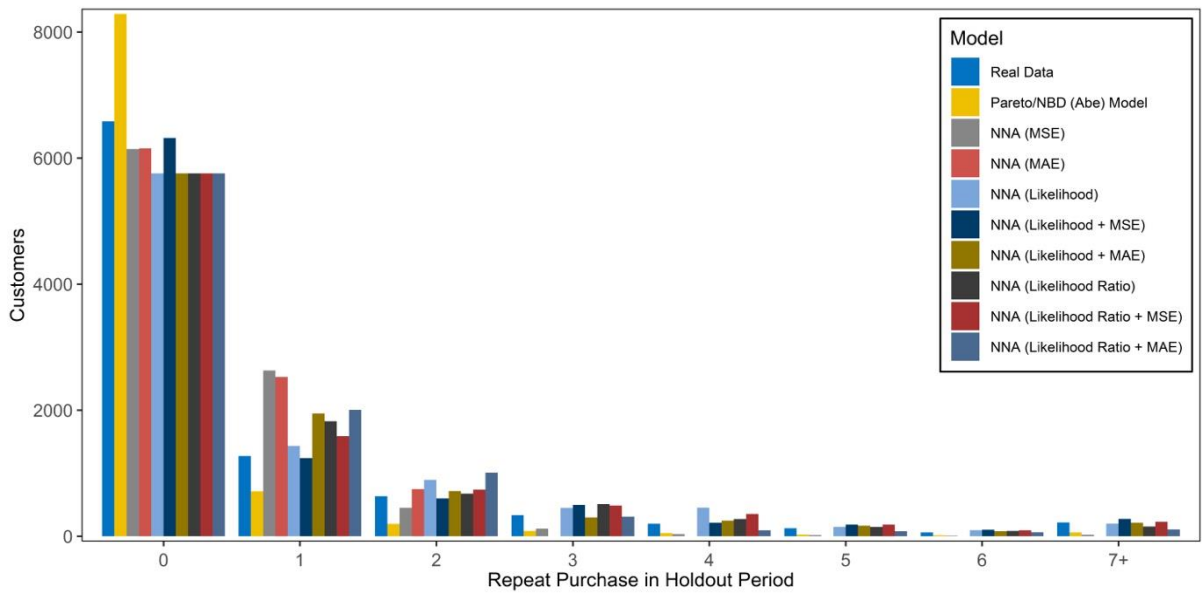

Figure 5. Predicted Versus Actual Customer Number of Repeat Purchases (CDNOW: With Covariate)

Checking the customer number behind Figure 5 in Table 9, NNA (Likelihood + MSE) has the best customer distribution over each repeat transaction point like that in the non-covariate setting. With the covariate effect, the Pareto/NBD (Abe) model pours more predictive focus on zero repeat transactions than does the Pareto/NBD model in Table 3. NNA (MSE) and NNA (MAE) have more predictive focus on zero and one repeat purchase fittings, and they even improve the fitting on zero repeat transactions compared with that in Table 3. These visualized conclusions also need further evidence from the evaluation metrics in order to select the best modelling candidate in the covariate setting.

Table 9. Predicted Versus Actual Customer Number of Repeat Purchases (CDNOW: With Covariate)

| Repeat Purchase Number | 0 | 1 | 2 | 3 | 4 | 5 | 6 | 7+ |
|---|---|---|---|---|---|---|---|---|
| **Pareto/NBD (Abe) model** | 8,288 | 710 | 196 | 84 | 49 | 26 | 15 | 60 |
| **Real Data** | 6,582 | 1,274 | 635 | 336 | 197 | 126 | 60 | 218 |
| **NNA (MSE)** | 6,144 | 2,629 | 452 | 119 | 34 | 19 | 8 | 23 |
| **NNA (MAE)** | 6,154 | 2,527 | 747 | 0 | 0 | 0 | 0 | 0 |
| **NNA (Likelihood)** | 5,757 | 1,432 | 894 | 449 | 453 | 149 | 96 | 198 |
| **NNA (Likelihood + MSE)** | 6,320 | 1,237 | 599 | 497 | 213 | 183 | 103 | 276 |
| **NNA (Likelihood + MAE)** | 5,757 | 1,948 | 716 | 299 | 248 | 167 | 80 | 213 |
| **NNA (Likelihood Ratio)** | 5,757 | 1,825 | 676 | 510 | 275 | 147 | 85 | 153 |
| **NNA (Likelihood Ratio + MSE)** | 5,757 | 1,587 | 737 | 487 | 353 | 184 | 95 | 228 |
| **NNA (Likelihood Ratio + MAE)** | 5,757 | 2,006 | 1,009 | 312 | 94 | 79 | 63 | 108 |

Compared to the non-covariate setting, the included covariates have a faint lifting effect on the inactive status forecasting. The Pareto/NBD (Abe) model, NNA (MAE), and NNA (MSE) obtain little improvement in identifying inactive customers.

Just like that of the Pareto/NBD model, the Pareto/NBD (Abe) model has an insufficient estimation in total repeat transaction prediction during the holdout period, where 7,634 transactions are made by 9,428 customers. For NNA (MAE) and NNA (MSE), the fitting improvement of customer number on zero repeat purchases has a negative influence on total purchase prediction. However, the models with a likelihood function in the loss function emit better modelling ability when the covariate effect is embedded in the model. NNA (Likelihood Ratio) shows the best fit at aggregate-level repeat purchase forecasting. NNA (Likelihood + MAE) and NNA (Likelihood + MSE) perform good as well. This result demonstrates that the additional constraints contribute to error computation, which can be passed on to the backpropagation process in order to obtain a better neural network structure.

Table 10. Evaluation Metrics (CDNOW: With Covariate)

| Model | Inactive Accuracy | Multi-accuracy | MAE | Consistency (NNA, ABE) | Number of Purchases |
|---|---|---|---|---|---|
| **Pareto/NBD (Abe) model** | 77.33% | 69.21% | 0.691 | - | 2,435 |
| **NNA (MSE)** | 77.12% | 62.20% | 0.729 | 72.76% | 4,367 |
| **NNA (MAE)** | 77.23% | 62.25% | 0.728 | 71.21% | 4,021 |
| **NNA (Likelihood)** | 76.47% | 56.42% | 0.929 | 61.99% | 9,615 |
| **NNA (Likelihood + MSE)** | 76.56% | 60.72% | 0.902 | 67.88% | 8,676 |
| **NNA (Likelihood + MAE)** | 76.45% | 57.17% | 0.878 | 62.12% | 8,690 |
| **NNA (Likelihood Ratio)** | 76.32% | 57.09% | 0.852 | 62.11% | 8,508 |
| **NNA (Likelihood Ratio + MSE)** | 76.54% | 56.49% | 0.916 | 62.04% | 9,497 |
| **NNA (Likelihood Ratio + MAE)** | 76.51% | 57.66% | 0.794 | 62.97% | 6,961 |

NNA (Likelihood + MSE) has the best proportional fitting on customer number at each repeat purchase point in Table 9, and there are similar evaluation metrics to NNA (Likelihood MAE) and NNA (Likelihood Ratio) in Table 10. Moreover, the consistency between the neural network algorithms and Pareto/NBD (Abe) model shows evidence that higher consistency brings a greater total repeat purchase estimation. In general, this study concludes that NNA (Likelihood + MSE) is the optimal modelling candidate in the covariate setting.

Table 11. Correlations Between Multi-accuracy, MAE, Number of Purchases, and Consistency (CDNOW: With Covariate)

| Correlation | Multi-accuracy | MAE | Number of Purchases | Consistency |
|---|---|---|---|---|
| **Multi-accuracy** | 1.0000 | | | |
| **MAE** | -0.8011 | 1.0000 | | |
| **Number of Purchases** | -0.8894 | 0.9800 | 1.0000 | |
| **Consistency** | 0.9852 | -0.7464 | -0.8529 | 1.0000 |

Same as with the correlations in Table 5, the signs between the multi-accuracy, MAE, and the number of purchases remain unchanged. This indicates that the predictive weight on zero or small repeat purchases contributes to small aggregate-level repeat transactions and higher multi-accuracy, which also lead to greater MAE. Additionally, the consistency pattern remains unchanged. The superior predictability of the neural network algorithms on aggregate-level repeat transaction fitting derives from the inconsistent individual-level repeat purchase prediction to the Pareto/NBD (Abe) model. Hence, this demonstrates that data analysts cannot just focus or rely on evaluation metrics as the evaluation standard to select the optimum modelling candidate.

2. E-tailing dataset

When a covariate is added, the customer number fitting on each repeat purchase point shows prompt improvement, especially for NNA (MAE) and NNA (MSE) at zero repeat purchase fitting. However, abnormal fitting of these two models exists in the remaining repeat purchase points. Moreover, the Pareto/NBD (Abe) model does not fit well for high-frequent customers in the holdout period. Just like the above analysis, Table 12 examines the customer number behind each repeat purchase point in Figure 6.

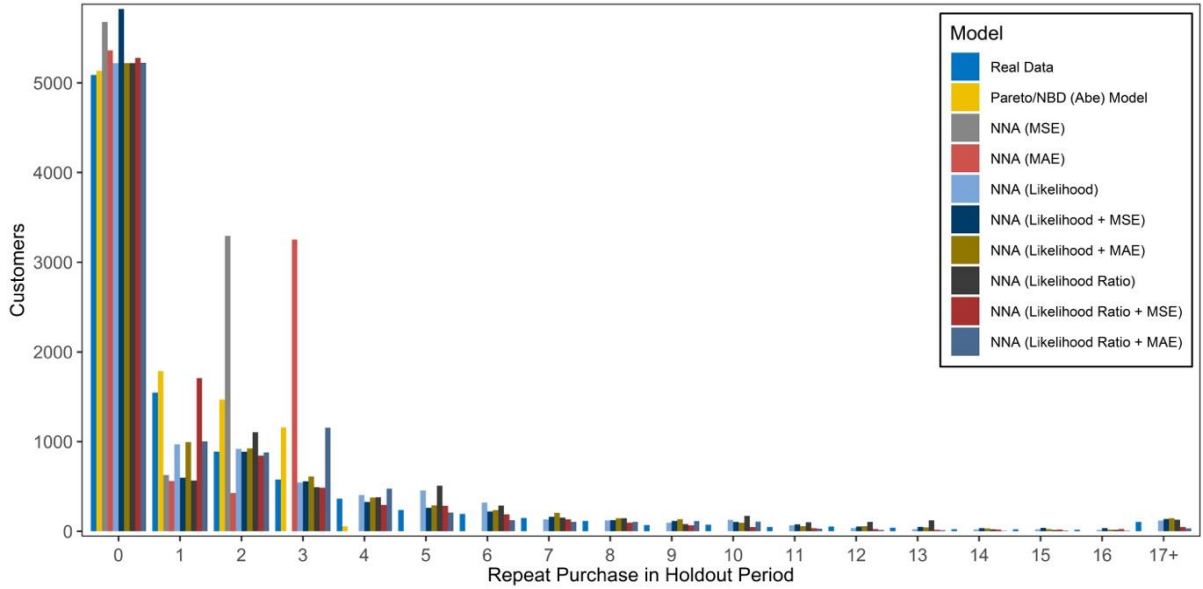

Figure 6. Predicted Versus Actual Customer Number of Repeat Purchases (E-tailing: With Covariate)

Table 12 reports the customer distribution under the with-covariate setting. The results suggest that the Pareto/NBD (Abe) model, NNA (MAE), and NNA (MSE) cannot be the modelling candidate due to shrinkage in prediction after the covariate is included. By eye-balling, the neural network algorithm with likelihood function obtains a better approximation to the true customer distribution. NNA (Likelihood Ratio + MSE) shows the best customer distribution approximation. Next, the evaluation metrics in Table 13 are utilized to select the best model.

Table 12. Predicted Versus Actual Customer Number of Repeat Purchases (E-tailing: With Covariate)

| Repeat Purchase Number | 0 | 1 | 2 | 3 | 4 | 5 | 6 | 7 | 8 | 9 | 10 | 11 | 12 | 13 | 14 | 15 | 16 | 17+ |
|---|---|---|---|---|---|---|---|---|---|---|---|---|---|---|---|---|---|---|
| Pareto/NBD (Abe) Model | 5,135 | 1,785 | 1,467 | 1,158 | 54 | 1 | 0 | 0 | 0 | 0 | 0 | 0 | 0 | 0 | 0 | 0 | 0 | 0 |
| Real Data | 5,087 | 1,545 | 889 | 577 | 363 | 238 | 194 | 147 | 115 | 69 | 72 | 48 | 52 | 39 | 23 | 21 | 17 | 104 |
| NNA (MSE) | 5,680 | 627 | 3,293 | 0 | 0 | 0 | 0 | 0 | 0 | 0 | 0 | 0 | 0 | 0 | 0 | 0 | 0 | 0 |
| NNA (MAE) | 5,362 | 561 | 426 | 3,251 | 0 | 0 | 0 | 0 | 0 | 0 | 0 | 0 | 0 | 0 | 0 | 0 | 0 | 0 |
| NNA (Likelihood) | 5,221 | 969 | 920 | 544 | 404 | 455 | 319 | 131 | 120 | 94 | 127 | 65 | 35 | 23 | 20 | 21 | 14 | 118 |
| NNA (Likelihood + MSE) | 5,825 | 598 | 888 | 557 | 327 | 262 | 221 | 160 | 123 | 115 | 103 | 77 | 52 | 49 | 35 | 38 | 34 | 136 |
| NNA (Likelihood + MAE) | 5,221 | 994 | 924 | 611 | 375 | 289 | 237 | 206 | 144 | 134 | 93 | 56 | 56 | 43 | 31 | 24 | 18 | 144 |
| NNA (Likelihood Ratio) | 5,221 | 566 | 1,104 | 490 | 379 | 507 | 286 | 151 | 145 | 79 | 170 | 99 | 104 | 121 | 21 | 14 | 15 | 128 |
| NNA (Likelihood Ratio + MSE) | 5,278 | 1,707 | 843 | 485 | 294 | 284 | 190 | 131 | 96 | 65 | 46 | 33 | 24 | 15 | 19 | 19 | 24 | 47 |
| NNA (Likelihood Ratio + MAE) | 5,224 | 1,002 | 881 | 1,154 | 474 | 210 | 123 | 104 | 104 | 113 | 107 | 28 | 14 | 12 | 5 | 7 | 6 | 32 |

In Table 13 the neural network algorithms with likelihood function have the same capacity in identifying inactive customers as do the other models. NNA (MSE) and NNA (MAE) can also be used for inactive customer identification, but not for repeat purchase forecasting, because of the severe deviation from total repeat transactions in the holdout period (17,159 transactions). The Pareto/NBD (Abe) model has an insufficient repeat purchase estimation, even with a comparable multi-accuracy and MAE to the other models. However, these comparable evaluation metrics are from the conservative estimation at individual purchase fitting, which appears in Table 12. Among the models, NNA (Likelihood + MSE) has the best aggregate-level repeat purchase prediction, presenting good multi-accuracy and MAE. NNA (Likelihood) could be the suboptimal modelling choice, because it has similar metrics to NNA (Likelihood + MSE). All the other models have severe deviation from the true total repeat transactions in the holdout period. Hence, no matter how similar multi-accuracy and MAE these models have, they cannot be taken into modelling consideration. In conclusion, NNA (Likelihood + MSE) is the modelling candidate in the covariate setting.

Table 13. Evaluation Metrics (E-tailing: With Covariate)

| Model | Inactive Accuracy | Multi-accuracy | MAE | Consistency (NNA, ABE) | Number of Purchases |
|---|---|---|---|---|---|
| **Pareto/NBD (Abe) model** | 67.99% | 44.59% | 1.504 | - | 8,414 |
| **NNA (MSE)** | 68.00% | 46.61% | 1.562 | 65.20% | 7,213 |
| **NNA (MAE)** | 69.67% | 44.11% | 1.610 | 62.78% | 11,166 |
| **NNA (Likelihood)** | 68.10% | 42.93% | 1.687 | 52.00% | 18,847 |
| **NNA (Likelihood + MSE)** | 68.74% | 46.43% | 1.663 | 60.79% | 18,345 |
| **NNA (Likelihood + MAE)** | 68.19% | 42.91% | 1.711 | 52.32% | 19,899 |
| **NNA (Likelihood Ratio)** | 68.60% | 42.21% | 1.836 | 51.17% | 21,982 |
| **NNA (Likelihood Ratio + MSE)** | 68.74% | 44.98% | 1.427 | 57.40% | 13,920 |
| **NNA (Likelihood Ratio + MAE)** | 69.24% | 43.69% | 1.501 | 56.10% | 15,054 |

Table 14. Correlation Between Multi-accuracy, MAE, and Number of Purchases (E-tailing: With Covariate)

| Correlation | Multi-accuracy | MAE | Number of Purchases | Consistency |
|---|---|---|---|---|
| **Multi-accuracy** | 1.0000 | | | |
| **MAE** | -0.5150 | 1.0000 | | |
| **Number of Purchase** | -0.6141 | 0.7176 | 1.0000 | |
| **Consistency** | 0.8609 | -0.4964 | -0.8648 | 1.0000 |

Table 14 lists the same patterns as Table 5 and Table 11, but has some reverse patterns to Table 8. The negative correlations between the multi-accuracy and MAE or the number of purchases demonstrate that higher exactly forecasting is unable to bring about better individual-level repeat purchase estimation. When comparing to that in Table 8, there is a weaker correlation between the number of purchases and MAE when the covariate effect is introduced. All these differences show that the covariate effect has a significant impact on repeat purchase forecasting in this dataset. Furthermore, the relative consistency with the Pareto/NBD (Abe) model returns a better repeat purchase prediction. Again, data analysts should be concerned about the trade-off between the metrics and the predicted individual quantity.